\documentclass[doublecol]{epl2} 

\usepackage{graphicx} 
\usepackage{subfig}
\usepackage{amssymb}

\usepackage[english]{babel}
\usepackage{color} 
\newcommand{\bq}{\begin{equation}} 
\newcommand{\eq}{\end{equation}}
\newcommand{\ba}{\begin{eqnarray}} 
\newcommand{\ea}{\end{eqnarray}}

\usepackage{comment}

\begin{document}
\title{Parasites on parasites: coupled fluctuations in stacked contact processes}
\shorttitle{Parasites on parasites}
\author{Steven J.~Court \and Richard A.~Blythe \and Rosalind J.~Allen}
\shortauthor{S.~J.~Court \etal}

\institute{SUPA, School of Physics and Astronomy, University of Edinburgh, Mayfield Road, Edinburgh EH9 3JZ, UK}


\abstract{
We present a model for host-parasite dynamics which incorporates both vertical and horizontal transmission as well as spatial structure. 
Our model consists of stacked contact processes (CP), where the dynamics of the host is a simple CP on a lattice while the dynamics of the 
parasite is a secondary CP which sits on top of the host-occupied sites. In the simplest case, where infection does not incur any cost, we 
uncover a novel effect: a nonmonotonic dependence of parasite prevalence on host turnover. Inspired by natural examples of hyperparasitism, 
we extend our model to multiple levels of parasites and identify a transition between the maintenance of a finite and infinite number of 
levels, which we conjecture is connected to a roughening transition in models of surface-growth.
}
\maketitle 

\section{Introduction}

The need to understand and control the dynamics of infections has motivated the development of a variety of statistical mechanical models. One of 
the most important of these is the contact process (CP) \cite{CPHarris, ob:Marro, ob:Hinrichsen, ob:Liggett}, which describes the dynamics of an infection 
in a spatially structured population. 
In the contact process, 
each site on a lattice represents a host organism which can be \emph{infected} or \emph{susceptible} (uninfected). Infection is transmitted to neighbouring susceptible host  sites at rate $b$, and infected hosts recover (\emph{i.e.}~become susceptible) at rate $d$. The contact process provides a baseline model for many problems in  ecology and epidemiology. It is also  
of fundamental importance  in statistical physics. This is because it exhibits a non-equilibrium phase transition between an infected and a non-infected 
phase at a critical value of $\lambda=b/d$ \cite{ob:Hinrichsen, ob:Marro}. 

The standard CP model for an infected population assumes that the  host population is of fixed size, without turnover, \emph{i.e.}~host births and deaths. This is valid if
the timescale on which the infection is gained and lost is much shorter than the lifespan of an individual. Some infections, however, are
carried by individuals for long times, and may be transmitted to offspring upon reproduction (vertical transmission), as well as being transmitted
horizontally---\emph{i.e.} upon physical contact between individuals. Examples include plasmids carried by bacteria \cite{plasmid_book},
some microsporidian parasites of insects \cite{Microsporidians} and 
pathogens including HIV and several hepatitis viruses. Here, we investigate the interplay between vertical and horizontal transmission in such populations. Other 
authors have determined the conditions for parasite persistence in mean-field models that lack spatial structure, in which the parasite may affect host 
fitness \cite{Lipsitch95,JonesVertical,BusenbergVertical}, and have suggested that these conditions may be affected by spatial structure  \cite{Schinazi}.  Here, we 
show that spatial structure can  produce a qualitatively new effect: a coupling between the  dynamics of the infection and  of the underlying host population, even 
when the infection does not affect the fitness of the host. 

We present a \emph{two level stacked contact process}, in which the host population is represented by a CP on a lattice, and the parasite population is 
represented by a second CP which sits on top of the host CP. This model incorporates both vertical and horizontal transmission --- if an infected host 
reproduces or dies, the 
parasite is reproduced or dies with it (vertical transmission), and a parasite-infected site can infect a neighbouring site if it is occupied by an uninfected 
host (horizontal transmission).  We characterize the conditions for parasite persistence in the form of a phase diagram and discover an interesting phenomenon: 
although the   steady-state 
properties of the host population depend only on the ratio of its birth and death rates, $\lambda=b/d$, the prevalence of the parasite can depend non-monotonically 
on the host population's turnover rate. This phenomenon has its origins in the fluctuations of spatial clusters of host 
individuals, and cannot straightforwardly be captured by a mean-field theory.

Parasitic infections  are not always limited to two levels.  \emph{Hyperparasitism}, in which  an organism carrying a primary parasite is susceptible 
to a secondary parasite \cite{HyperparasiteReview}, can be harnessed as a biocontrol mechanism --- examples include viral infections of the 
fungus {\em{Cryphonectria parasitica}} that causes chestnut blight \cite{Milgroom2004}, and cytoplasmic RNA elements that infect the fungus causing 
Dutch elm disease \cite{Swinton1999}. 
Inspired by these scenarios, we extend our model to a {\em{multilevel stacked contact process}}, in which 
individuals carrying a primary infection may be susceptible to secondary infections, and those carrying the secondary infection may be susceptible 
to tertiary infections, etc. 
This raises a number of questions: How does the dynamics of one parasite level couple to the next, and how many 
levels of parasites are sustainable in a population? 
We show that in our stacked contact process, there is a well-defined transition between maintenance of an infinite hierarchy of levels of parasites, and limitation 
to a finite number of levels.  The transition between these two regimes appears to be connected to a roughening transition in 
certain surface growth processes \cite{Alon1996,Alon1998,Blythe2001}.  This work presents
new challenges to our understanding of contact processes, with potential implications for the dynamics of long-timescale infections.

\section{Two-level stacked contact process}

\label{sec:model}

\begin{figure}[t]
  \begin{center}
  \includegraphics[width=6cm]{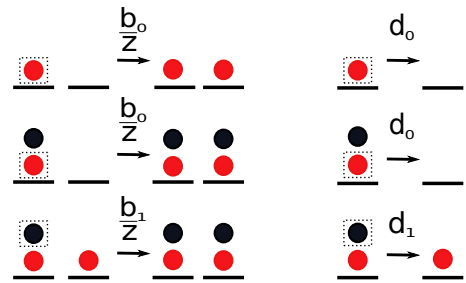}
  \end{center}
  \caption{The events that constitute the 2-level stacked CP. Red and black circles represent host organisms and parasites respectively. 
A site can be empty, occupied by a susceptible host  or occupied by an infected host. 
The left panels represent reproduction of a susceptible or infected host (top and middle, rate $b_0$), and transmission of the parasite 
(bottom, rate $b_1$). The right panels represent death of   a susceptible or infected host (top and middle, rate $d_0$), and loss of the 
parasite (bottom, rate $d_1$).  $z$ is the number of nearest neighbour sites. The dotted squares indicate the level of the CP at which the event happens. }
  \label{fig:Dynamics1}
\end{figure}

We begin by considering the dynamics of a host and a single parasite which can be transmitted vertically or horizontally --- a two-level stacked CP --- on a 2D square lattice. Lattice sites can be either empty or occupied 
by a host organism. Occupied lattice sites become empty at rate $d_0$, due to death of hosts, and host organisms attempt to replicate into neighbouring empty sites at rate $b_0$ 
(the attempt rate in any direction is $b_0/z$, $z$ being the coordination number of the lattice). Host organisms may 
also carry a parasite, which we assume to be neutral in the sense that organisms with and without the parasite have identical birth and death rates. 
Infected individuals pass the parasite to uninfected neighbours at rate $b_1$, and the parasite is lost at a constant rate $d_1$. We assume that the offspring of an infected organism is also infected (vertical transmission), and that when an infected host dies, 
the parasite dies with it. 
Thus the infection dynamics consists of a secondary CP, with parameters $b_1$ 
and $d_1$, which sits on top of the host's birth-death CP. Since the host population does not occupy the whole lattice, 
the secondary CP does not take place on a regular lattice but rather on the irregular network of occupied sites which changes stochastically 
due to the dynamics of the host population. 
The processes that occur in our model are illustrated in Figure \ref{fig:Dynamics1}. We simulate this model using a 
stochastic kinetic Monte Carlo algorithm, in which all events occur as Poisson processes \cite{BKL}, on a $100\times100$ 
lattice with periodic boundary conditions. Averages are calculated after an initial transient to allow the steady state to be reached.

\section{Host dynamics influences parasite persistence}\label{sec:2level}
In this model, the host population undergoes a simple CP, whose stationary properties are fully determined by the single 
parameter $\lambda_0 = b_0/d_0$ \cite{CPHarris, ob:Marro, ob:Hinrichsen, ob:Liggett}. If $\lambda_0$ is less than a critical value, $\lambda_{\mathrm{crit}}$, the only steady state 
is an empty lattice and the population rapidly becomes extinct. If $\lambda_0 > \lambda_{\mathrm{crit}}$, a non-zero population can be 
maintained for long times, before eventually becoming extinct due to a rare fluctuation. The transition between these two regimes is 
second order; the values of $\lambda_{\mathrm{crit}}$  and the associated critical
exponents have been characterised in detail \cite{ob:Hinrichsen}. Defining $\tau_i = 0$ or $1$ if site $i$ is empty or occupied 
respectively, the dynamics of the host population obeys
\begin{eqnarray}\label{eq:cp}
 \frac{{\rm d}}{{\rm d}t}\left<\tau_i\right> = \frac{1}{z}\sum_j b_0 \left<\tau_j\left(1-\tau_i\right)\right> - d_0 \left<\tau_i\right> \;,
\end{eqnarray}
where the sum is over the $z$ neighbours of lattice site $i$ and the angle brackets denote
averages over multiple realizations of the dynamics. The steady-state solution of Eq.~(\ref{eq:cp}) satisfies
$\left<\tau_i\right>  = \left(\lambda_0 / (\lambda_0-1)\right)\left<\tau_i \tau_j\right> $. Ignoring correlations between neighbouring sites by
assuming that $\left<\tau_i \tau_j\right>  \approx \left<\tau_i\right>\left<\tau_j\right>$, we arrive at the mean-field result
$\left<\tau\right> = 1-(1/\lambda_0)$.

Considering now the dynamics of the parasite, we define $\sigma_i = 1$ if site $i$ contains an infected host and $\sigma_i = 0$ otherwise
(\emph{i.e.} if site $i$ is either unoccupied or contains a susceptible host).  $\sigma_i$ obeys
\begin{eqnarray}\label{eq:sigma}
\nonumber  \frac{{\rm d}}{{\rm d}t}\left<\sigma_i\right> &=& \frac{1}{z}\sum_j \left[ b_0 \left< \sigma_j\left(1-\tau_i\right)\right> + b_1\left<\sigma_j\left(\tau_i-\sigma_i\right)\right>\right]\\      &&{} - d_0\left<\sigma_i\right> - d_1\left<\sigma_i\right> \;.
\end{eqnarray}
The first term on the r.h.s.~of Eq.~(\ref{eq:sigma}) corresponds to vertical transmission: an empty site is filled by replication of an infected host. The second term represents horizontal transmission: a susceptible host, denoted by $\left(\tau_i-\sigma_i\right)$,
is infected by a parasite-carrying neighbour. The final two terms correspond to death of an infected host and loss of the parasite. Eq.~(\ref{eq:sigma}) can be rewritten as
\begin{eqnarray}\label{eq:sigma2}
\nonumber\frac{{\rm d}}{{\rm d}t}\left<\sigma_i\right> &=&  \frac{1}{z}\sum_j \left[ b_0 \left< \sigma_j \right> + (b_1- b_0)\left< \tau_i\sigma_j\right> - b_1 \left<\sigma_i\sigma_j \right> \right]\\ 
                                  && {}- (d_1+d_0)\left<\sigma_i\right> \;.
\end{eqnarray}
Comparing this with Eq.~(\ref{eq:cp}), we see that the cross-correlation
$\left< \tau_i\sigma_j\right>$ perturbs the parasite dynamics  from that of a standard CP. Interestingly, however, if $b_1=b_0$   (\emph{i.e.} if the rates of horizontal and vertical transmission are equal), this term vanishes and the form of Eq.~(\ref{eq:sigma2}) becomes that of a standard CP   with
parameter $\lambda_{\mathrm{eff}} = \lambda_0\lambda_1/(\lambda_0 + \lambda_1) = b/\sum_k d_k$. The parasite
dynamics also has CP-like behaviour in the   mean-field limit, where we neglect spatial correlations: setting $\left<\tau_i \sigma_j\right>  \approx \left<\tau_i\right>\left<\sigma_j\right>$ and
$\left<\sigma_i \sigma_j\right>  \approx \left<\sigma_i\right>\left<\sigma_j\right>$ in Eq.~(\ref{eq:sigma2}), we obtain the steady-state solution
$\left< \sigma \right> = 1- (1/\lambda_0) - (1/\lambda_1) \equiv 1-(1/\lambda_{\mathrm{eff}})$.

\begin{figure}[!t]
  \centering
  \includegraphics[width=5.5cm]{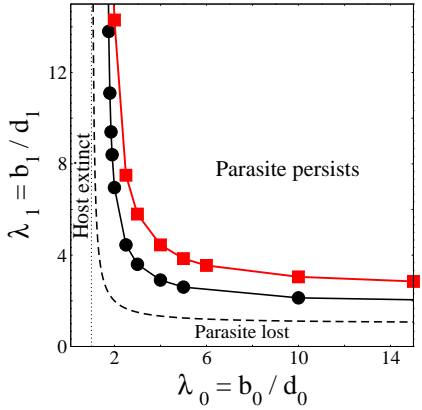}
  \caption{Phase diagram of the two-level stacked CP.  Symbols: simulation data showing boundary of parasite persistence, 
for $d_1=1$ and $d_0=1$ (circles) or $d_0=20$ (squares). Dashed line: mean-field prediction $\lambda_1 = \lambda_0/(\lambda_0-1)$.}
  \label{fig:PhaseDiagram}
\end{figure}

Figure \ref{fig:PhaseDiagram} shows the phase diagram for our model, as a function of $\lambda_0$ and $\lambda_1$. Three steady state scenarios are
possible: (i) the host population is extinct, (ii) the host population is finite but the parasite is extinct, and (iii) the parasite persists within
a finite host population. Since the dynamics of the host population is a standard CP, the host population is extinct (scenario (i)) if
$\lambda_0 < \lambda_{\mathrm{crit}}$. For $\lambda_0 > \lambda_{\mathrm{crit}}$, the host population persists, with or without the parasite.  The condition for parasite persistence (\emph{i.e.} the boundary separating  scenarios (ii) and (iii))  is predicted by the mean field theory to be  $\lambda_1 = \lambda_0/(\lambda_0-1)$
(obtained by setting $\left< \sigma \right> = 0$) -- shown by the dashed line in Figure  \ref{fig:PhaseDiagram}. As one might expect, as  $\lambda_0$ decreases, the density of the host population decreases, and a higher rate of horizontal transmission ($\lambda_1$) is needed to maintain the parasite. 

The symbols in Figure \ref{fig:PhaseDiagram} show the boundaries between parasite persistence and loss obtained from our kinetic Monte Carlo simulations, for two values of $d_0$, the host death rate. The fact that these are shifted upwards and to the right of  the mean-field prediction shows that spatial correlations make 
it harder to maintain the parasite. It is also interesting that we obtain different results for the two different values of $d_0$. This shows that the parameters $\lambda_0$ and $\lambda_1$ do not fully determine 
the phase behaviour of the system. For fixed 
$\lambda_0=b_0/d_0$, a higher turnover rate of the host population (\emph{i.e.} higher $d_0$ and $b_0$) apparently makes it harder to maintain the parasite.

\begin{figure}[!b]
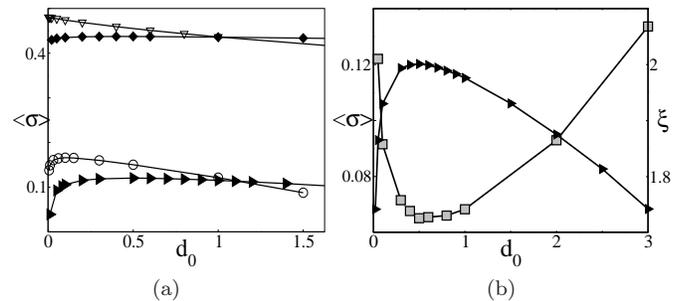

  \centering
  \subfloat[]{\label{fig:VaryingSpeeds}\includegraphics[width=0.48\linewidth, height=3.45cm]{Fig3a_noLegend}}\hfill
  \subfloat[]{\label{fig:Correlation}\includegraphics[width=0.51\linewidth]{Fig3b_CorrLength}}
  \caption{(a) Parasite density $\left<\sigma\right>$ as a function of host turnover rate $d_0$ for fixed $\lambda_0$ (and hence fixed host density). Curves top to bottom correspond to $(\lambda_0,b_1,d_1)=(5,4,1)$,$(2.5,15,1)$,$(2,8,1)$ and $(1.8,15,1)$. (b) Parasite density $\left<\sigma\right>$ (triangles, left axis) and spatial correlation length $\xi$ (squares, right axis) for $(\lambda_0,b_1,d_1)=(1.8,15,1)$.}
\end{figure}

In fact the situation is more complex. Figure  \ref{fig:VaryingSpeeds} shows a more comprehensive investigation of the steady-state parasite 
density $\left<\sigma\right>$ as a function of the host turnover rate. In these simulations, we vary $b_0$ and $d_0$ keeping $\lambda_0 = b_0/d_0$ fixed. The 
steady-state  density of the host population is constant, since it depends only on  $\lambda_0$. Remarkably, for some parameter combinations,  the density of the 
parasite actually depends non-monotonically on the host turnover rate. If the host dynamics is slow, increasing the turnover rate increases the parasite density, 
while if the host dynamics is fast, the parasite density decreases with the turnover rate.

This dependence of the parasite density on the host turnover rate is absent in the mean-field theory and must therefore be a consequence of spatial 
correlations. To investigate this, we measure in our simulations the spatial correlations of the parasite density via the function 
$C_{i,j}=\left[\langle \sigma_i \sigma_j \rangle / \langle \sigma \rangle\right] - \langle \sigma\rangle$.
Here, the first term is the conditional probability of finding a parasite at site $j$ given that there is one at site $i$, while the second term 
ensures that  $C_{i,j} \rightarrow 0$ as $|i-j| \rightarrow \infty$ (where $|i-j|$ denotes the distance between sites $i$ and $j$). We then 
fit our data to the functional form $C_{i,j}\sim {\rm e}^{-|i-j|/\xi}$ to extract the spatial correlation length $\xi$, which can be thought 
of as the parasite cluster size. This is shown in Figure \ref{fig:Correlation}. Strikingly, the  correlation length is minimal at the turnover 
rate where the parasite density is maximal. This makes intuitive sense: since horizontal transmission requires contact between infected and 
uninfected hosts, minimal spatial  clustering of the parasite population results in maximal parasite density. 

Why should the turnover rate of the host affect the spatial clustering of the parasite? At fast host turnover rates, we observe in our simulations 
that local patches of host organisms rapidly grow from a single seed and  vanish by stochastic extinction, on a faster timescale than that of 
horizontal transmission. Because offspring are always of the same type as their parents, organisms in a single patch are either all infected or all uninfected. 
Thus the infected population is highly clustered, leading to low rates of horizontal transmission and low parasite density. 
By contrast, when the host population turnover is slow, the parasite dynamics constitutes a CP on the effectively frozen, disordered network of lattice sites that are occupied by host organisms. This network contains clusters of sites that may be disconnected, or poorly connected, from the rest of the host population; this creates the possibility of local, stochastic parasite extinctions. 
This is most obviously the case in the low $\lambda_0$ regime where the host density is low, but it is still true for higher host densities.  In this regime, an increase in the host turnover rate serves as a mixing mechanism, by more homogeneously distributing the host 
population and thereby providing greater opportunity for the parasite to spread.

Thus the nonmonotonic dependence of the parasite density on the host turnover rate arises from a coupling of the spatial fluctuations of the underlying host CP to the parasite dynamics, and 
 can be viewed as a competition between the mixing effect of the host birth-death process 
at low turnover rates and the population segregation arising from parent-offspring clustering at high turnover rates.

\section{Multi-level stacked contact processes}

\begin{figure}[!t]
  \centering
  \includegraphics[width=6cm]{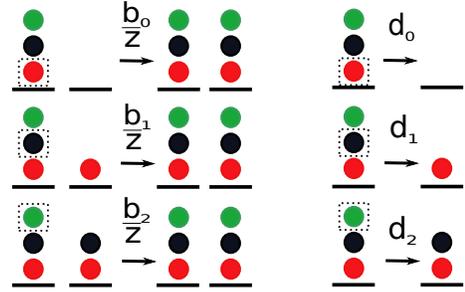}
  \caption{Possible transitions for the multi-level CP, considering a site with label $m=3$. The symbols are as in Fig.~\ref{fig:Dynamics1}; the green circles represent the secondary parasite. Birth processes are shown on the left. The organism in question can reproduce into a neighbouring empty site (top left), transmit its primary parasite to a susceptible neighbour (middle left - note the secondary parasite is also transmitted), or transmit its secondary parasite to a 
susceptible neighbour (bottom left). Death processes are shown on the right. These consist of death of the host (top right), loss of the primary parasite (middle right) or loss of the secondary parasite (bottom right). }
  \label{fig:Dynamics2}
\end{figure}

Motivated by natural examples of hyperparasitism, we now investigate a {\emph{multi-level}} stacked CP,  where individuals carrying a 
primary infection are susceptible to a secondary infection, and those with the 
secondary infection are susceptible to tertiary infections, and so on.
In the $M$-level stacked CP, a site is labelled $0$ if it is empty, $1$ if it contains an uninfected host, and $m=2, 3, \ldots, M$ if it contains a host infected with $m-1$ 
levels of parasites. For example, a site labelled 2 contains a host infected with only a primary infection (parasite) while a site labelled 3 contains a host which is infected 
with primary and secondary infections (a parasite and a hyperparasite). Note that the presence of an infection at level $m$ implies the presence of lower-level 
infections (\emph{e.g.} one cannot have a secondary infection without a primary infection). Figure \ref{fig:Dynamics2} illustrates the 
dynamical processes that can occur in this model. An empty site may be occupied at rate $b_0/z$ by reproduction of a neighbouring site of any label (top left), or a 
site labelled $m$ may be infected at rate $b_m/z$ by higher-level parasites from a neighbouring site with label $n > m$; its label is then promoted to $n$ (middle and bottom left).  
The host organism of a level $m$ site can die, at rate $d_0$ (top right), or experience loss of one of its parasites, at rate $d_{n}$ (where $n \le m$ relates to the level 
of parasite that is lost). When the latter happens, all higher parasites are also lost and  the site is demoted to level $n$ (middle and bottom right panels in 
Fig.~\ref{fig:Dynamics2}).  
The parameters of the model are the  number of levels $M$, and the set of level-dependent 
birth and death rates $\{ b_0, \ldots, b_{M-1}; d_0, \ldots, d_{M-1}\}$.

We first consider the special case where the birth rates at all levels are equal; $b_n=b$ $\forall$ $n$. We divide lattice sites into two sets: 
sites that have labels greater than or equal to $m$ (here denoted ${\cal M}_{+}$), and sites that have labels less than $m$ (here denoted ${\cal M}_{-}$). An ${\cal M}_{-}$ site can 
become ${\cal M}_{+}$ by infection by a neighbouring ${\cal M}_{+}$ site: this occurs at rate $b/z$ regardless of the level of the ${\cal M}_{+}$ site. An 
${\cal M}_{+}$ site can become ${\cal M}_{-}$ by death of the host or by loss of a parasite at levels $n=2 \dots m$. This occurs at total rate $d=\sum_{\ell=0}^{m-1} d_\ell$
where the sum is over levels up to $m-1$.
Thus the dynamics of the ${\cal M}_{+}$ density is exactly equivalent to that of a  standard CP with 
parameter $\lambda_{\mathrm{eff}} = b/\sum_{\ell=0}^{m-1} d_\ell$.

\begin{figure}[t]
  \centering
  \includegraphics[width=5.25cm]{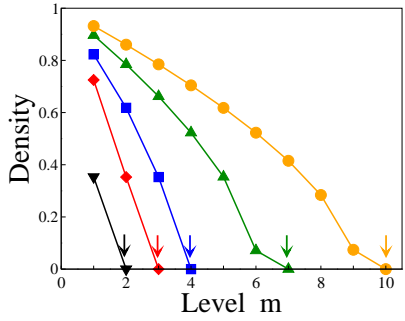}
  \caption{ Density of sites with label $m$ or higher, as a function of level $m$, for a stacked CP with equal birth rates $b_n=b$ and 
death rates $d_n=1$. Left to right correspond to simulations with $b=2$,$4$,$6$,$10$ and $15$ respectively. Arrows 
indicate $m^*=\lceil b / (\lambda_{\mathrm{crit}}d ) \rceil$, the boundary for sustainability as predicted by mapping to the standard CP.}
  \label{fig:SustainableLevels}
\end{figure}

This observation provides us with important insight into the number of levels that are sustainable in a stacked CP. 
For the standard (one-level) CP, a finite density of occupied lattice sites can only be sustained for $\lambda > \lambda_{\mathrm{crit}}$, where 
$\lambda_{\mathrm{crit}} \approx 1.649$ for a 2D square lattice \cite{ob:Hinrichsen}. In the stacked CP with equal birth rates
we therefore expect the density of ${\cal M}_{+}$ sites, those with label $n \ge m$, to be non-zero only if $\lambda_{\mathrm{eff}} = b_0 / \sum_{\ell=0}^{m-1}d_\ell  < \lambda_{\mathrm{crit}}$.
Figure \ref{fig:SustainableLevels} shows simulation results for the average density of lattice sites with label $n \ge m$, as a function of $m$, for a stacked CP with 
equal birth rates $b_n=b$ 
and death rates $d_n=d=1$, 
for several values of $b$. By mapping onto a standard CP  with $\lambda_{\mathrm{eff}} = b/(md)$, we predict that all levels $n \ge m^* = \lceil b / (\lambda_{\rm crit} d) \rceil$ have zero density and are thus unsustainable.  Our simulation results, shown in Figure \ref{fig:SustainableLevels}, bear this out: the system indeed only sustains a finite number $m^*$ of parasite levels.

\begin{figure}[t]
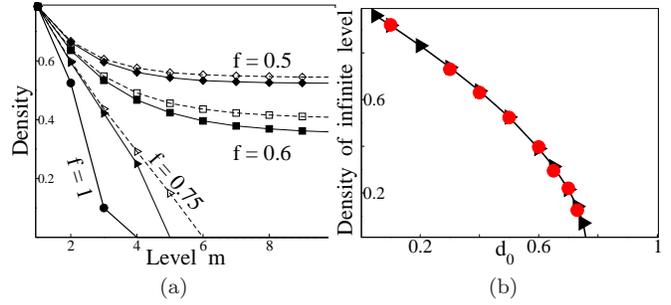

  \centering
  \subfloat[]{\label{fig:Vary_b_and_d}\includegraphics[width=0.49\linewidth]{Density_of_levels_BLACK}}
  \subfloat[]{\label{fig:Dens_inf}\includegraphics[width=0.49\linewidth]{Density_of_infinite_level2_noLegend} }
  \caption{(a) Density of sites with $n \ge m$, as a function of level $m$, in the stacked CP for decreasing death rates $d_n = f^n d_0$, with $d_0=1, b_n=5$ 
(solid lines) and for increasing birth rates $b_n = f^{-n}b_0$, with $b_0=5, d_n=1$ (dashed lines).  
In the former case, the exact CP predicts a nonzero density as $m\to\infty$ only for $f\gtrapprox 0.67$. 
(b) Density of the m$^{th}$ level in the limit $m\to\infty$ with $b_n=5, d_n=f^{n}d_0, f=0.75$ as a function of host death rate $d_0$ from simulations 
(circles) and as predicted by a standard CP (triangles) with $\lambda_{\mathrm{eff}}^{m\rightarrow \infty}=b(1-f)/d_0$.  The simulation data is obtained by measuring 
the plateau values in plots like panel (a), for $m$ values up to 20. The mapping to a standard 
CP predicts that only a finite number of levels are sustainable for  $d_0 > d_0^\ast \approx 0.758$.
 }
\end{figure}

Are there any circumstances where a stacked CP can be sustained for an infinite number of levels?
This is indeed possible if either the birth rate $b_n$ increases, or the death rate $d_n$ decreases, sufficiently strongly with $n$.  We first suppose that the birth rate $b_n=b$ is constant but the death rate decreases by a factor $f$ at successive levels: $d_n=fd_{n-1}=f^{n}d_0$ where $0 < f < 1$. In this case, the average density of sites with label $n \ge m$ is given by that of a standard CP with
$\lambda_{\mathrm{eff}} = b/\sum_{\ell=0}^{m-1} d_\ell = b(1-f)/\left[d_0(1-f^{m})\right]$. To sustain  an infinite number of levels, we require that
$\lambda_{\mathrm{eff}}^{m \rightarrow \infty} = b(1-f)/d_0 \ge \lambda_{\mathrm{crit}}$, which implies that 
$f \le 1 - (d_0\lambda_{\mathrm{crit}}/b)$. Our simulation results, Fig.~\ref{fig:Vary_b_and_d}, show that, with $d_0=1$ and $b=5$, the system indeed sustains an infinite number of levels for values of $f$ above the critical value of $f\approx0.67$.

Interestingly, this analysis also shows that, for given values of $b$ and $f$, there exists a critical value of the host death rate $d_0$ which
separates regimes where the stacked CP can and cannot sustain an infinite number of levels. This is given by $d_0^* = b(1-f)/\lambda_{\mathrm{crit}}$.
Figure \ref{fig:Dens_inf} shows, as a function of $d_0$, the predicted density of sites with label $n \ge m$, as  $m \to \infty$: this is given by
the density of a standard CP with $\lambda_{\mathrm{eff}} = b(1-f)/d_0$. The density as $m \to \infty$ indeed falls to zero at $d_0=d_0^*$. 

Finally, we explore briefly the case where, instead of  varying the death rate between levels, we instead increase the birth rate 
by a factor $1/f$ (\emph{i.e.} set $b_n = b_{n-1}/f$). In this case the dynamics no longer maps onto that of a standard CP -- but nevertheless, our 
simulations show qualitatively similar results (dashed lines in Figure \ref{fig:Vary_b_and_d}). 

The transition to an infinite number of levels of parasites that we observe in this model appears to be related to a 
transition that occurs in models for the growth of interfaces by deposition of particles on surfaces \cite{Alon1996,Alon1998,Blythe2001}. In these models, 
the interface is modelled as a lattice, with a given height at each lattice site. According to the  ballistic deposition rule for surface growth \cite{Meakin1986}, 
particles fall onto the lattice from above and only stick if the  neighbouring lattice site  already contains a particle. In some models, particles can also desorb 
from the surface  \cite{Blythe2001}. In the stacked CP model, we can think of the label $m$ of a given lattice site as corresponding to the local height of the interface. 
The transmission of higher-level parasites to a site with label $m$, from a neighbour with $n > m$ then corresponds to  ballistic deposition,
while death of a host and loss of parasites loosely correspond to desorption of particles. 
This apparent mapping is intriguing because these models for surface growth show a roughening transition \cite{Alon1996,Alon1998,Blythe2001}: when the deposition  rate is low, the interface remains \emph{smooth}, that is, the width of the surface layer remains finite in the thermodynamic limit, whereas 
when the deposition rate is high, the surface layer grows and \emph{roughens} over time, that is, arbitrarily large differences in surface height can arise in the thermodynamic limit. These two cases correspond to finite and infinite  hierarchies of parasites in our stacked CP model. 
Although the mapping to the surface growth models is not exact, one expects to see the same phenomenology, \emph{e.g.}~in terms of critical exponents, for the transitions. 
For example a generic model for coupled directed percolation processes with unidirectional coupling between adjacent levels \cite{Tauber1998} can show
a different $\beta$ exponent at different levels when the critical points coincide. 
It would be interesting to see if this is true for models such as the one described here which show bidirectional coupling between all levels.

\section{Discussion}
In this paper, we have presented a simple model for  the dynamics of long-lived infections in spatially structured populations.
The model is a stacked contact process (CP), 
in which the host population undergoes a standard CP, and a secondary CP representing the parasite takes place on the dynamically changing network of lattice sites occupied by the host. We find that the coupling 
between the dynamics of the host population and of the infection leads to non-trivial effects, including a non-monotonic dependence of the parasite prevalence on the host population's turnover rate, for fixed host population size. This exposes a connection between
spatial and temporal scales, since host population turnover affects spatial clustering of infected organisms which in turn affects the prevalence of the 
infection within the population. We have found that an  improved mean-field theory that takes local correlations into account 
(the ordinary pair approximation \cite{Sato94}) fails to reproduce this nonmonotonic behaviour. This  suggests that a
deep understanding of the fluctuations in the host CP is required to predict the steady-state properties of the parasite 
population. It will be interesting in future to investigate  how the nature of the CP phase transition for the infection dynamics is affected  by the dynamics of the underlying host population. Since disorder can alter critical exponents and lead to new dynamics such as Griffiths phases
and activated scaling \cite{GriffithPhase, Vojta2005}, we envisage that the secondary CP may show fundamentally different behaviour to the standard CP.

Inspired by natural examples of hyperparasitism, or ``parasites on parasites'', we have also investigated the properties of a {\emph{multi-level}} contact process.  
For this model, we show that the average density of the $m$-th (and above) level of the stacked CP maps onto that of a standard, one-level CP, if the birth rates at all levels are 
equal. We find that a phase transition separates two very different behaviours of the model: maintenance of an infinite hierarchy of levels of parasites, versus collapse to zero density at a finite number of levels. In light of the link to models of surface growth 
it will be interesting  to determine more fully  the properties of the parasite maintenance transition that we observe in the multi-layer stacked CP.

Finally, we note that the stacked CP model also presents the possibility for a novel study of disorder in contact processes. Classic
models for disordered systems usually create disorder by removing a fraction of sites from the dynamics \cite{MoreiraDickman1996} or by 
allowing different creation or annihilation rates for different sites \cite{NoestDisorder}. This disorder is often quenched - \emph{i.e.} time-invariant. 
In our model, disorder in the secondary CP (and higher CPs if present) arises naturally from
the underlying dynamics of the primary CP, and can be controlled by varying the rates. This naturally arising disorder is not fixed but varies in time and space (in
a fundamentally different process than the mobile disorder of Ref.~\cite{DickmanMobile}). 
Future studies of the effects of this natural disorder on the properties of the secondary CP may 
uncover new principles of CP dynamics on disordered lattices.

\section{Acknowledgements}
We thank Martin Evans for useful discussions. SJC was funded by the Carnegie Trust for the Universities of Scotland, 
RAB by an RCUK Academic Fellowship and RJA by a Royal Society University Research Fellowship.

\bibliographystyle{eplbib}

\end{document}